\documentclass[letterpaper]{article} 
\usepackage{aaai2026}  
\usepackage{times}  
\usepackage{helvet}  
\usepackage{courier}  
\usepackage[hyphens]{url}  
\usepackage{graphicx} 
\usepackage{natbib}  
\usepackage{caption} 
\usepackage{algorithm}
\usepackage{algorithmic}
\usepackage{amsmath} 
\usepackage{amssymb}
\usepackage{amsfonts}

\usepackage{soul} %
\usepackage{color, xcolor} 
\sethlcolor{yellow} %

\usepackage{newfloat}
\usepackage{listings}
\DeclareCaptionStyle{ruled}{labelfont=normalfont,labelsep=colon,strut=off} 
\floatstyle{ruled}
\newfloat{listing}{tb}{lst}{}
\floatname{listing}{Listing}
\usepackage{xspace}
\usepackage{xcolor}
\usepackage{multirow}
\newcommand{\eat}[1]{}
\newcommand{\stitle}[1]{\textbf{#1}}

\newcommand{\ie}{\emph{i.e.},\xspace}

\newcommand{\modelname}{AutoDW\xspace}  
\newcommand{\benchname}{DWBench\xspace}

\title{Automating Complex Document Workflows via Stepwise and Rollback-Enabled Operation Orchestration}

\author {
    Yanbin Zhang,
    Hanhui Ye,
    Yue Bai,
    Qiming Zhang,
    Liao Xiang,\\
    Wu Mianzhi,
    Renjun Hu\thanks{Corresponding author}
}
\affiliations {
    East China Normal University\\
    ybzhang@dase.ecnu.edu.cn, 
    51265903057@stu.ecnu.edu.cn, 
    baiyue@cc.ecnu.edu.cn, 
    qmzhang@stu.ecnu.edu.cn, \\
    51285903046@stu.ecnu.edu.cn, 
    mianzhiwu@stu.ecnu.edu.cn, 
    rjhu@dase.ecnu.edu.cn

}

\usepackage{bibentry}
\usepackage{float}

\begin{document}

\maketitle

\begin{abstract}
Workflow automation promises substantial productivity gains in everyday document-related tasks. While prior agentic systems can execute isolated instructions, they struggle with automating multi-step, session-level workflows due to limited control over the operational process. To this end, we introduce \textbf{\modelname}, a novel execution framework that enables stepwise, rollback-enabled operation orchestration. \modelname incrementally plans API actions conditioned on user instructions, intent-filtered API candidates, and the evolving states of the document. It further employs robust rollback mechanisms at both the argument and API levels, enabling dynamic correction and fault tolerance. These designs together ensure that the execution trajectory of \modelname remains aligned with user intent and document context across long-horizon workflows. To assess its effectiveness, we construct a comprehensive benchmark of 250 sessions and 1,708 human-annotated instructions, reflecting realistic document processing scenarios with interdependent instructions. \modelname achieves 90\% and 62\% completion rates on instruction- and session-level tasks, respectively, outperforming strong baselines by 40\% and 76\%. Moreover, \modelname also remains robust for the decision of backbone LLMs and on tasks with varying difficulty. Code and data will be open-sourced.
\begin{links}
\link{Code}{https://github.com/YJett/AutoDW}
\end{links}
\end{abstract}


\section{Introduction}
\label{sec:intro}

Automating document-related tasks, which constitute a substantial portion of the intellectual workload, remains a persist challenge.With the advent of large language models (LLMs)~\cite{brown2020languagemodelsfewshotlearners}, significant progress has been made in language understanding, instruction following, and multi-step planning~\cite{comanici2025gemini25pushingfrontier,guo2025deepseek}. This has opened new possibilities for workflow automation that involve interpreting instructions, maintaining context, and translating intent into structured actions. Explorations in code generation~\cite{ishibashi2024self}, data science~\cite{hong-etal-2025-data}, and web-based tasks~\cite{yang2025agentoccam} have demonstrated encouraging results. These advances highlight the promise of extending LLM-powered automation to complex document workflows for productivity improvement.


Several lines of research could apply for this purpose. 
General tool use or workflow orchestration models~\cite{schick2023toolformer,fan2024workflowllm} have been developed to facilitate effective API selection. However, they are not tailored for precise document state management, especially in long-horizon tasks. 
Other document-oriented agents adopt a human-in-the-loop paradigm, sacrificing automation for alignment~\cite{mathur-etal-2024-docpilot,liang2025tabletalk}. 
More closely related to our setting are recent agentic systems designed for instruction-based automation~\cite{guo2023pptcbenchmarkevaluatinglarge}. Leveraging the native planning ability of LLMs~\cite{yao2023react,rawat2025preact}, these systems have demonstrated promising performance in isolated instruction execution. \textbf{Nonetheless}, such agents typically employ predetermined plans without context-sensitive adjustments, struggling to precisely align flexible and, sometimes, ambiguous user instructions with evolving document states. \textbf{Moreover}, they generally lack error recovery mechanisms, \ie directly proceeding without validating outcomes. Consequently, even a single erroneous API call can propagate silently, limiting effectiveness  in managing session-level multi-step workflows~\cite{wang2024officebenchbenchmarkinglanguageagents, yao2023docxchainpowerfulopensourcetoolchain}.

These challenges underscore the necessity of more structured and precisely controlled operation orchestration for automating complex document workflows. To address this, we introduce $\text{\modelname}$, a novel execution framework that provides a general and robust orchestration paradigm through the integration of stepwise planning and adaptive rollback. Unlike previous methods that treat the task as a monolithic process, \modelname incrementally decomposes workflows into atomic operations. Each operation is executed and verified individually, allowing errors or misalignments with user intent to be detected early and corrected promptly.

Specifically, \modelname incrementally plans API actions one at a time, conditioning each choice on the user instruction, intent-filtered candidate APIs, and the latest document state. 
Within a Python runtime environment, \modelname executes the selected API and extracts the updated document state. 
It then invokes an LLM to summarizes the state change before and after execution. 
In the adaptive rollback stage, an LLM-based validator assesses the alignment of the state change with user intent, providing a confidence score and detailed explanation. Based on this feedback, \modelname chooses to accept the API or employs two distinct rollback strategies: argument-level rollback, which revises only the API arguments while preserving the API selection, and API-level rollback, which completely re-generates an alternative API. By integrating precise incremental planning with adaptive rollback, \modelname effectively maintains alignment with user intent throughout complex sessions, significantly reducing the risk of cascading errors.


To rigorously evaluate this robust orchestration capability, we construct a new benchmark \textbf{\benchname} designed to reflect realistic document processing scenarios involving highly interdependent and long-horizon instructions.
\benchname consists of 250 multi-turn sessions (1,708 instructions total). To support these workflows, we implement 74 APIs and hired graduate-level volunteers to annotate each instruction with a feasible API sequence. Notably, most instructions require multiple API calls, with each session averaging 34.8 APIs ($\text{min 15}$, $\text{max 75}$), demonstrating the non-triviality and depth of \benchname  for automation.


%
For evaluation, we adopt an operational correctness metric inspired by recent agent benchmarks~\cite{zhou2024webarena,yao2025taubench}.Specifically, we ask an LLM-based judge to compare the programatically-extracted document state after each executed instruction against the ground-truth state obtained by replaying the annotated API sequence. A task is considered successfully completed if the LLM determines the two states are semantically equivalent.To ensure reliability, we conduct a specialized analysis to verify the judge's high agreement with human verification on a sampled subset of tasks.

With \benchname, we find that \modelname achieves 90\% instruction-level and 62\% session-level completion rates, outperforming strong baselines by 40\% and 76\%, respectively. The latter improvement is achieved by using an extra of 25.6\% tokens, demonstrating significant gains in end-to-end workflow completion. Our robustness study also verifies that $\text{\modelname}$ generalizes well across different backbone $\text{LLMs}$ (DeepSeek-v3, Qwen-Plus, Gemini-2.5-Pro, and GPT-4.1) and varying instruction difficulty. Moreover, the performance on hard instructions (\ie those require $> 6$ $\text{APIs}$ to fulfill) is only 4.4\% lower than the overall
suggesting its practical applicability in real-world usage scenarios. Finally, our ablation study convincingly confirms the rationale of our rollback strategies, which could strike a good balance between performance gain and extra cost with single-round dual-level rollback. 

In summary, this paper makes the following contributions:
\begin{itemize}
    \item We introduce \modelname, which integrates an innovative combination of stepwise planning and adaptive rollback for complex document workflow automation.
    \item We construct \benchname for evaluation, which features 250 sessions of 1,708 human-annotated instructions, 74 APIs, and  an operational correctness metric.
    \item With \benchname, we empirically verify that \modelname obtains the state-of-the-art performance the the task, as well as its robustness and rationale in model design.
\end{itemize}

\section{Related Work}
\label{sec:related}

\stitle{LLM-based document workflow automation}.
Recent advances in LLMs have enabled new possibilities for automating document-related tasks through natural language interfaces. Early work demonstrated their potential for basic text understanding and generation~\cite{brown2020languagemodelsfewshotlearners}, but subsequent evaluations revealed limitations in reliably executing multi-step document operations~\cite{bang2023multitask,borji2023categorical}. Specialized architectures (e.g., LayoutLM~\cite{xu2020layoutlm}, DocLLM~\cite{wang2024docllm}) have advanced document understanding, but focus on comprehension rather than execution reliability.

General $\text{LLM}$-based agents (e.g., $\text{ReAct}$~\cite{yao2023react}, $\text{Toolformer}$~\cite{schick2023toolformer}) provide effective frameworks for planning and tool use, but are not designed for the document domain.

Existing document automation approaches typically fall into three paradigms. 
Retrieval-based methods rely on semantic similarity to select relevant APIs~\cite{karpukhin2020dense}, but they often misalign with user intent due to the gap between natural language and function semantics. 
Reasoning-only approaches employ LLMs to generate executable actions directly~\cite{wei2023chainofthoughtpromptingelicitsreasoning}, offering flexibility but suffering from brittleness and inconsistency in long-horizon workflows. 
Hybrid approaches~\cite{guo2023pptcbenchmarkevaluatinglarge,mathur-etal-2024-docpilot} combine planning, selection, verification, and execution, but their predetermined planning lacks adaptability and error recovery is generally neglected.
In contrast, \modelname directly addresses these challenges with state-aware stepwise planning and rollback mechanisms.

\begin{figure*}[tb]
\centering
\includegraphics[width=0.95\textwidth]{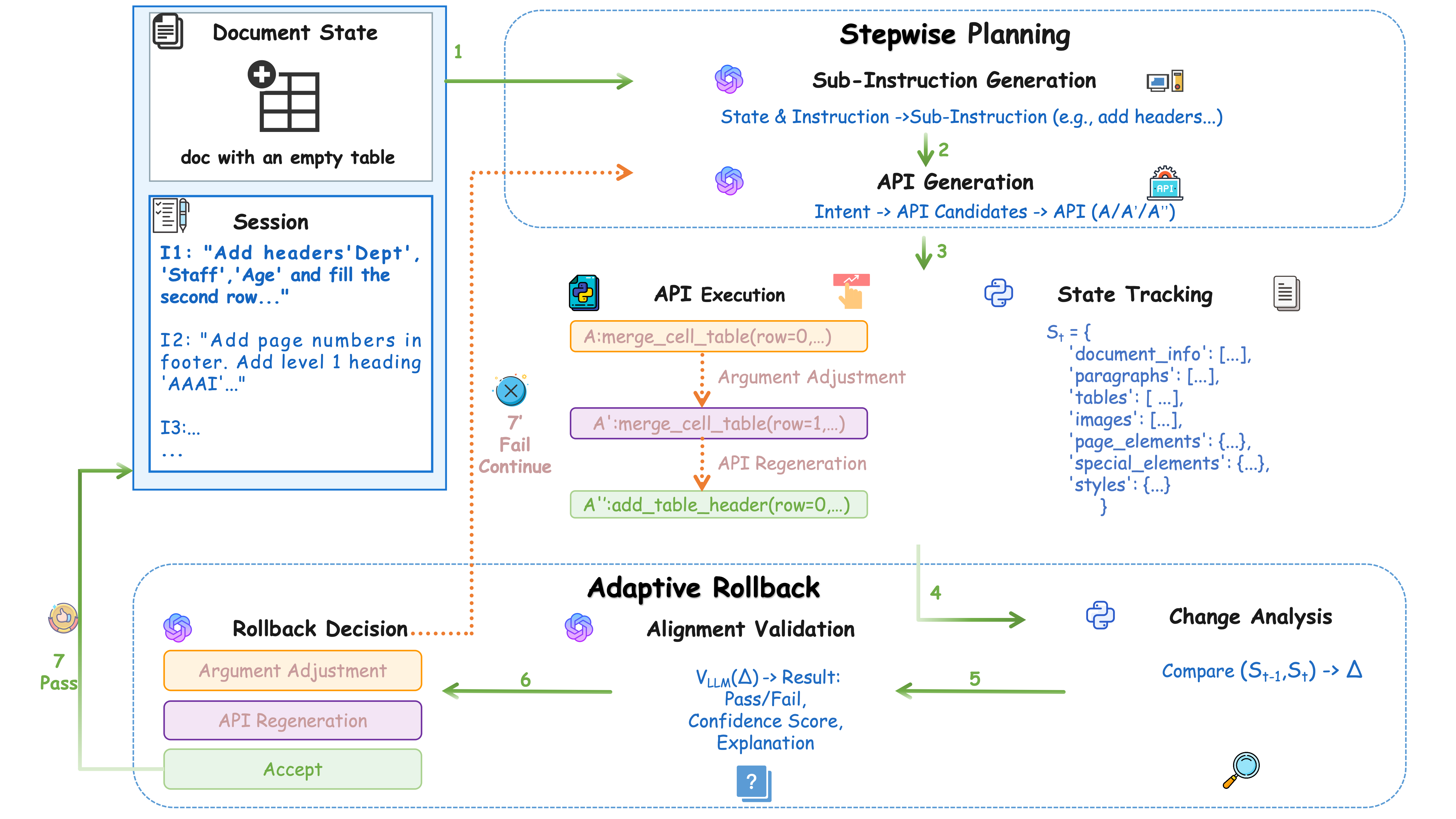}
\caption{Overview of the \modelname framework, which comprises three core modules: stepwise planning, API execution \& state tracking, and adaptive rollback. The overview also includes an illustrative example showing how \modelname selects one API call at a time and corrects its mistakes (\ie APIs A and A') through rollback.}
\label{fig:system_framework}
\end{figure*}

\stitle{Execution reliability and fault recovery in agents}.
Reliability is a critical requirement for automation systems, especially in high-stakes or productivity-focused environments~\cite{vendrow2025_reliability_benchmarks}. Traditional error handling strategies have evolved into self-healing~\cite{academy28_tor_metric} and multi-level validation~\cite{academia30_llm_kb_reliability}. However, these domain-agnostic systems struggle with document-specific concerns where subtle changes propagate unintended effects.

Efforts to improve robustness in sequential tasks include stepwise execution in fields like program synthesis~\cite{wang2024incremental} and robot planning~\cite{paxton2019prospection}. These techniques operate in deterministic environments, which are not directly transferable to document workflow automation characterized by natural language ambiguity inherent in natural language instructions. \modelname extends these ideas by introducing intent-conditioned validation and rollback that accommodate natural language variability.

\stitle{Document automation benchmarking}.
Despite growing interest in document automation, evaluation frameworks have lagged behind in assessing real-world task reliability. Previous benchmarks emphasize static analysis tasks such as information extraction~\cite{mathew2021docvqadatasetvqadocument}. DocBench~\cite{zou2024docbenchbenchmarkevaluatingllmbased} focuses on document analysis but does not evaluate execution reliability. 
$\text{PPTC}$~\cite{guo2023pptcbenchmarkevaluatinglarge} introduced a benchmark for multi-turn PowerPoint automation, where $\text{GPT-4}$ achieved only 6\% session completion. Similarly, we construct $\text{\benchname}$, a bilingual benchmark for $\text{Word}$ automation (250 workflows), featuring human-annotated $\text{API}$ sequences and operational correctness evaluation, offering a realistic measure for production-grade agents.

Each instruction is annotated by human experts with a feasible API sequence, and it adopts operational correct for evaluation.
Together, these features make \benchname a more realistic measure of readiness for production-grade document automation agentic systems.

\section{Methodology}
\label{sec:method}

In this section, we present our proposed framework, \modelname, which performs stepwise and rollback-enabled operation orchestration for complex document workflow automation. An overview of the system architecture is shown in Fig.~\ref{fig:system_framework}
Given an initial document state $\mathcal{S}_0$ (\ie a Word file) and a session of natural language instructions $\mathcal{I}=\{I_1, I_2, \dots\}$, \modelname plans and executes a sequence of API calls $\mathcal{P}=[A_1, A_2, \dots]$ that incrementally transform $\mathcal{S}_0$ in alignment with the user intent expressed in $\mathcal{I}$. Each API $A_i$ is selected from a predefined API set $\mathcal{A}$, which \modelname has full access to.
At a high level, \modelname comprises three core modules:
a stepwise planning module for incrementally selecting the next API,
a Python runtime environment for executing APIs and tracking document states, and
an adaptive rollback module for validating and potentially reverting previously executed operations.
We now describe each of these components in detail.


\subsection{Stepwise Planning}

Traditional instruction decomposition approaches generate all atomic operations upfront, which often leads to execution failures as the document state evolves. In contrast, our stepwise planning mitigates this limitation by generating API calls one at a time, conditioned on real-time document state, ensuring that each operation adapts to the current execution context. While this is a locally guided approach, the comprehensive and up-to-date document state encoding (as detailed in Table~\ref{tab:doc_state}) provides sufficient context, effectively preventing the system from falling into low-efficiency or dead-end local optima during complex, long-horizon sessions.
Specifically, we adopt a two-stage planning strategy: first generating a sub-instruction, and then the corresponding API action.

\stitle{Sub-instruction generation}.
Each sub-instruction corresponds to a single atomic document operation that can be completed with one API call. The use of sub-instructions serves two key purposes. First, it bridges the semantic gap between potentially ambiguous natural language instructions and concrete API functionalities. Second, it enables intent classification, which narrows down the API search space during API generation by identifying the likely user intent behind the atomic operation.
Sub-instruction generation is performed by an LLM prompted with a comprehensive template that includes the task objective, the original user instruction, previously completed API calls, the current document state (as captured in the state tracking module), detailed guidelines, output format specifications in JSON, and in-context examples.\footnote{All prompt templates used by \modelname are provided in the supplementary material.}

\begin{table*}[t]
\centering
\begin{tabular}{l|l}
\hline
\textbf{Component Type} & \textbf{Tracking Attributes} \\ \hline
Document Info & total paragraphs count, total tables count, total sections count, has header flag, has footer flag \\ 
Paragraph Elements & index, text content, style name, text alignment, text runs$^1$, spacing, indentation, embedded images$^2$ \\ 
Table Elements & index, row count, column count, cell matrix, table style, row heights, column widths \\ 
Image Elements & host paragraph index, host text run index, image sequence index, width, height \\ 
Page Layout Elements & headers, footers, page numbers, watermarks, table of contents \\ 
Interactive Elements & hyperlinks, bookmarks, line breaks, page breaks \\ 
Document Styles & style name, style category, font name, font size, bold flag, italic flag \\ \hline
\end{tabular}
\caption{Document state tracking details. $^1$Text runs are formatted text segments within a paragraph (e.g., ``normal \textbf{bold} normal" contains 3 runs). $^2$Embedded images are positioned within paragraph text runs, not standalone image objects.}
\label{tab:doc_state}
\end{table*}

\stitle{API generation}.
Once a sub-instruction is generated, \modelname classifies its underlying intent. Given that document-related intents are relatively fixed and well-defined, we fine-tune a 178M BERT model~\cite{devlin-etal-2019-bert} for 8-way intent classification: \emph{content creation}, \emph{content modification}, \emph{table operation}, \emph{image operation}, \emph{chart operations}, \emph{format/style editing}, \emph{document structure update}, and \emph{document lifecycle update}. The classifier is fine-tuned on 3,315 instruction–intent pairs (with no overlap with the \benchname benchmark), and achieves a test accuracy of 98\%.
This choice, instead of using the $\text{LLM}$ for classification, is primarily motivated by efficiency and cost: for a fixed and well-defined set of $\text{8}$ intents, $\text{BERT}$ maintains high accuracy while significantly reducing inference latency and computational overhead, optimizing the overall resource consumption of the framework.”

To improve robustness to ambiguous instructions, we retain the top-3 predicted intents instead of relying solely on the top-1. This allows \modelname to better capture the true intent behind diverse user inputs. The APIs associated with the top-3 detected intents are then retrieved and, along with the sub-instruction and other prompt engineering elements, fed into an LLM to generate the final API action, \ie an API and its full arguments.
Despite the framework's multi-stage reliance on $\text{LLMs}$, our design incorporates a subsequent robust validation and rollback mechanism that systematically mitigates the risk of $\text{LLM}$ errors accumulating and propagating across different modules in the execution chain.”

For further details on intent categories, associated APIs, and the BERT fine-tuning procedure, please refer to the supplementary material.




\subsection{API Execution and State Tracking}

\begin{equation}
\mathcal{S}_t = (\mathcal{D}^{\text{doc}}_t, \mathcal{D}^{\text{para}}_t, \mathcal{D}^{\text{table}}_t, \mathcal{D}^{\text{image}}_t, \mathcal{D}^{\text{page}}_t, \mathcal{D}^{\text{int}}_t, \mathcal{D}^{\text{style}}_t),
\end{equation}
where each $\mathcal{D}$ is a list of dict objects that describes all elements of a specific type in $\mathcal{S}_t$. We summarize the detailed attributes (\ie dict keys) to track for each type of components in Table~\ref{tab:doc_state}.

\eat{
The metadata $\mathcal{D}^{\text{meta}}_t$ includes structural counts where $n_{\text{para}} = |\text{doc.paragraphs}|$, $n_{\text{table}} = |\text{doc.tables}|$, and $h_{\text{header}}, h_{\text{footer}} \in \{0,1\}$. Each paragraph $p_i \in \mathcal{D}^{\text{para}}_t$ contains index, text, style, alignment, run-level formatting $\mathcal{R}_i$, spacing $\mathcal{S}_i$, and indentation $\mathcal{I}_i$. Run formatting $r_j \in \mathcal{R}_i$ captures text attributes including bold, italic, font, size, and color properties.

Table extraction captures structure through:
\begin{equation}
t_k = (\text{index}_k, \text{rows}_k, \text{cols}_k, \mathcal{C}_k, \text{style}_k, \text{signature}_k)
\end{equation}
where $\mathcal{C}_k$ represents the cell matrix. Page elements include headers, footers, and page numbers, while special elements encompass hyperlinks and bookmarks.

The extraction process uses modular extractors with shared utility classes. As shown in Algorithm~\ref{alg:state_extraction}, the complete extraction function combines specialized extractors:
\begin{equation}
\mathcal{S}_t = \mathcal{E}_{\text{complete}}(\text{doc}) = \bigoplus_{i \in \mathcal{I}} \mathcal{E}_i(\text{doc})
\end{equation}
where $\mathcal{I} = \{\text{meta}, \text{para}, \text{table}, \text{image}, \text{page}, \text{special}, \text{style}\}$.

}
Robustness against State Tracking Failures: We address this by treating a state parsing failure as an invalid execution outcome. If Python Runtime fails to extract a valid document state, the subsequent adaptive rollback module recognizes this as a critical misalignment with expected user intent, triggering an API-level rollback. This mechanism prevents the system from generating subsequent plans based on unreliable or erroneous document states.

\subsection{Adaptive Rollback}
\label{sec:Adaptive Rollback}

The rollback module in \modelname is responsible for handling fault recovery during execution. It achieves this by summarizing the document state change after an API call and validating whether the change aligns with the user’s intent. \modelname supports two types of rollback mechanisms: argument-level and API-level. The use of rollback is adaptively determined by an alignment validator.

\stitle{Change analysis}.
We implement an \texttt{analyze\_change} function which employs a multi-layered approach to detecting changes between document states. The analysis operates on state pairs $(\mathcal{S}_{t-1}, \mathcal{S}_{t})$ and outputs:
\begin{equation}
\Delta = (\Delta_S, \Delta_C, \Delta_F, \Delta_{St}, \Delta_T, \Delta_H),
\end{equation}
that represent structural, content, format, style, table, and hyperlink changes, respectively. 
These changes are captured by analyzing the seven-component document state: structural analyzer processes document info, paragraphs, tables, images, and page elements for element count changes; content analyzer examines paragraph and table text modifications; format analyzer handles run-level formatting within paragraphs and tables; style analyzer tracks paragraph style changes; table analyzer focuses on table-specific structural and content changes; and hyperlink analyzer processes special elements for link modifications.

The content analyzer uses Python's SequenceMatcher to detect text modifications, generating operation codes $\mathcal{O} = \{(\text{tag}, i_1, i_2, j_1, j_2)\}$ where $\text{tag} \in \{\text{insert}, \text{delete}, \text{replace}, \text{equal}\}$. The structural analyzer compares element counts and positions, while table analysis uses cell signature comparison:
\begin{equation}
\text{sig}_{i,j} = \text{hash}(\text{content} \| \text{structure} \| \text{position} \| \text{merge})
\end{equation}

\begin{figure*}[t]
\centering
\includegraphics[width=1.00\textwidth]{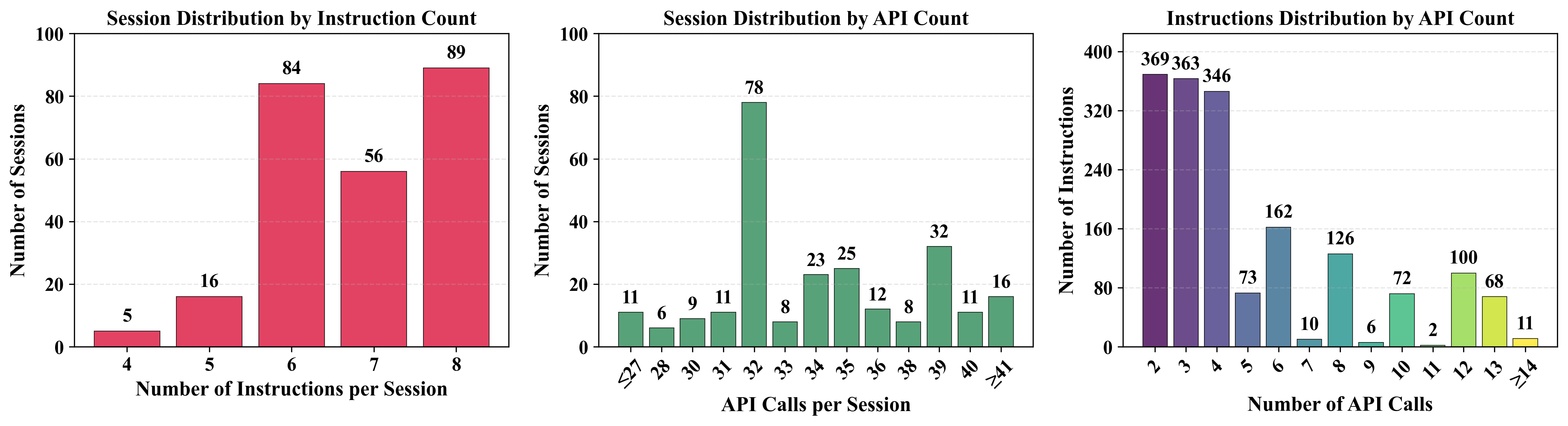}
\caption{Distributional statistics of \benchname, which includes 250 sessions and 1,708 instructions.
\textbf{Left}: Number of instructions per session (range: 4--8, mean=6.8), with a peak at 8 (89 sessions, 35.6\%).
\textbf{Middle}: Number of API calls per session (range: 15--75, mean=34.8), peak at 32 (78 sessions, 31.2\%).
\textbf{Right}: Number of API calls per instruction (range: 2--22, mean=5.1), while most instructions require 2-–4 API calls, complex instructions with $ge 10$ calls account for 14.8\%.}
\label{fig:api-distribution}
\end{figure*}

\stitle{Alignment validation and rollback}.
To perform validation, the system passes the analyzed state change $\Delta$, along with the sub-instruction, current document state, and previously executed APIs, to an LLM. The validator LLM is prompted to return:
(1) a binary decision (pass or fail),
(2) a confidence score in the range $[0, 1]$, and
(3) a textual explanation supporting its judgment.
Based on the result, \modelname decides whether to accept the action or initiate rollback. 
Specifically, the action is accepted if the LLM decision is 'pass' with high confidence (we use an empirically tuned threshold of 0.6), or if the confidence score is low. The selection of this threshold is supported by a sensitivity analysis in Ablation Study, demonstrating its optimal balance between false negatives and false positives.
In the latter case, low confidence is interpreted as the validator being uncertain, and the action is accepted by default. In all other cases, \modelname initiates a rollback to avoid potential execution errors.

Rollback follows a two-step routine. First, argument-level rollback is applied: \modelname regenerates the action using the same API but with updated arguments, taking validator’s explanation into consideration. If this revised action also fails validation, the system escalates to API-level rollback, where a new API is selected altogether. These two steps together form a full rollback round. 
While additional rollback rounds are computationally possible, $\text{AutoDW}$ defaults to a single round of rollback. This pragmatic choice is justified by our empirical analysis (see $\text{Section 4.4}$), which shows only marginal performance gains beyond the first attempt, making the single-round strategy the most cost-effective solution. 
With API-level rollback, regardless of the result, the most recent action is accepted and execution proceeds.

\stitle{Running example}. 
An example of AutoDW’s stepwise planning and self-correction through rollback is illustrated in Fig.~\ref{fig:system_framework}.
Given the instruction {\em ``Add headers `Dept', `Staff', `Age' and fill the second row ...''}, \modelname first decomposes the instruction into a sub-instruction {\em ``add headers ...''}. 
However, the initially generated API is \texttt{merge\_cell\_table(row=0,)}, which is incorrect. After executing this API, updating the document state, and analyzing the resulting state change, the alignment validator rejects the action due to misalignment with the intended operation. \modelname then initiates rollback. 
It first attempts argument-level rollback, generating a revised action with \texttt{row=1}, but this is still rejected. The system then escalates to API-level rollback, producing a new API: \texttt{add\_table\_header(row=0,)}, which is aligned with the sub-instruction and thus accepted. With this correction, \modelname resumes planning the next sub-instruction in the instruction.

\section{Experiments}
\label{sec:experiment}

We conduct an extensive evaluation of \modelname from three perspectives:
(1) a comparative analysis of overall performance against representative baselines,
(2) a robustness study examining performance across different backbone LLMs and tasks with varying difficulty, and
(3) an ablation study investigating the contributions of key components.


\subsection{Evaluation Setup}

We first introduce the experimental setting.

\stitle{Benchmarking data}.
To evaluate the effectiveness of \modelname, we construct a benchmark, \benchname, consisting of 250 sessions and 1,708 instructions. Each instruction is associated with: (i) a turn ID, (ii) a user-issued natural language instruction, (iii) an initial document state file, (iv) a human-annotated sequence of feasible API calls that fulfill the instruction, and (v) the corresponding expected document state obtained by executing the annotated API sequence on the initial state. The distributional statistics of \benchname are visualized in Fig.~\ref{fig:api-distribution}.


\stitle{Baseline approaches}. 
Document automation agents most relevant to our work include DocPilot~\cite{mathur-etal-2024-docpilot}, TableTalk~\cite{liang2025tabletalk}, and the systems evaluated in PPTC~\cite{guo2023pptcbenchmarkevaluatinglarge}. We exclude DocPilot and TableTalk from our comparison, as both rely on human-in-the-loop verification, which is incompatible with our fully automated evaluation setup. Adapting the evaluation protocol established in PPTC, we consider the following three baselines for comparison:

\noindent (1) Retrieval-only: This baseline performs semantic matching between the user instruction $I$ and the available APIs in $\mathcal{A}$, followed by naive execution using default arguments. First, both the instruction and each API description are embedded into dense vectors, and cosine similarity is computed. APIs with similarity above a threshold ($\tau=0.75$) are retained as candidates. These candidates are then ranked by similarity in descending order and instantiated using default argument values to form the final API call sequence.

\noindent (2) Reasoning-only: in this baseline, the full API library $\mathcal{A}$ is included in the LLM’s context along with user instruction $I$. The LLM is responsible for selecting appropriate APIs, determining their execution order, and generating fully parameterized API calls based on its own reasoning capabilities, without any external validation or decomposition steps.

\noindent (3) Hybrid: To enable a more thorough comparison, we implement the multi‐stage pipeline from PPTC~\cite{guo2023pptcbenchmarkevaluatinglarge}. At each turn $t$, the system receives the current instruction $I_t$, the dialogue history, the current document state, and a reference API list. An instruction-understanding module converts this input into an abstract intent representation $\mathcal{U}_t$. A rule-based mapper then selects and parameterizes the most relevant APIs to construct the API sequence $\mathcal{A}_t$. 


\eat{
\[
f_{\mathrm{retrieval}}:\;
\mathcal{I}
\;\longrightarrow\;
\mathcal{A}_{\mathrm{candidate}}
=\{A_i\in\mathcal{A}\mid \mathrm{sim}(\mathcal{I},A_i)\ge0.75\}.
\]
\(\mathcal{A}_{\mathrm{candidate}}\).

\[
f_{\mathrm{reasoning}}:\; \mathcal{I} + \mathcal{A}
\;\longrightarrow\;
\mathcal{D}_{\mathrm{modification}},
\]
where \(\mathcal{D}_{\mathrm{modification}}\) denotes the generated API call sequence.

\[
\begin{aligned}
f_{\mathrm{pptc}}:\;
&(I_t,\,\mathrm{history},\,\mathrm{ppt}_{\mathrm{state}},\,\mathrm{api}_{\mathrm{ref}}) \\
&\mkern-25mu\longrightarrow\;\mathcal{U}_t \longrightarrow\;\mathcal{A}_t \longrightarrow\;\mathrm{PPTX\text{-}Match} \longrightarrow\;\{\mathrm{success, fail}\}.
\end{aligned}
\]
}

\begin{table*}[h]
\centering
\setlength{\tabcolsep}{1.5mm}
\begin{tabular}{l|ccc|ccc}
\hline
\multirow{2}{*}{\textbf{Approach}} 
  & \multicolumn{3}{c|}{\textbf{Instruction-level}} 
  & \multicolumn{3}{c}{\textbf{Session-level}}  \\
& \textbf{iACC (\%)} & \textbf{\#APIs} & \textbf{\#Tokens (k)} & \textbf{sACC (\%)} & \textbf{\#APIs} & \textbf{\#Tokens}  \\ \hline
Retrieval-only & 13.84 & 4.82 & 29.6k & 4.40  & 15.50 & 98.7k  \\
Reasoning-only & 39.93 & 5.12 & 31.6k & 25.20 & 34.98 & 175.4k \\
Hybrid  & 64.46 & 5.30 & 36.5k & 35.20 & 36.21 & 225.2k  \\
\modelname (ours) & \textbf{90.33} & 5.21 & 42.8k & \textbf{62.00} & 34.45 & 284.9k \\ \hline
\modelname vs. Hybrid & +40.1\% & -1.7\% & +17.4\% & +76.1\% & -4.9\% & +26.5\% \\
\hline
\end{tabular}
\caption{Overall performance comparison with baselines. Tests were repeated for three times and we report the average.}  
\label{tab:baselines}
\end{table*}

\begin{table*}[h]
\centering
\setlength{\tabcolsep}{1.4mm}
\begin{tabular}{l|ccc|ccc|ccc}  
\hline
\multirow{2}{*}{\textbf{LLM}} 
& \multicolumn{3}{c|}{\textbf{Instruction-level}}  & \multicolumn{3}{c|}{\textbf{Session-level}} 
& \multicolumn{3}{c}{\textbf{iACC (\%) by difficulty}}  \\ 
& \textbf{iACC (\%)} & \textbf{\#APIs} & \textbf{\#Tokens} & \textbf{sACC (\%)} & \textbf{\#APIs} & \textbf{\#Tokens} & \textbf{S} & \textbf{M} & \textbf{H} \\ 
\hline      
Qwen-Plus
& 82.82 & 6.15 & 40.1k
& 53.60 & 38.60 & 265.4k
& \textbf{86.34} & \underline{83.13} & 78.99
\\
DeepSeek-v3
& 90.33 & 5.21 & 42.8k
& 62.00 & 34.45 & 284.9k
& \textbf{94.54} & \underline{90.02} & 86.33
\\
Gemini-2.5-Pro
& 91.79 & 4.92 & 42.7k
& \textbf{67.20} & 32.88 & 286.0k
& \textbf{95.08} & \underline{93.46} & 86.84
\\
GPT-4.1
& \textbf{92.03} & 5.14 & 43.5k
& 65.20 & 34.82 & 294.7k
& \textbf{95.77} & \underline{92.48} & 87.85
\\
\hline
\end{tabular}
\caption{Robustness study of \modelname in terms of backbone LLMs and task difficulty. We classify instruction into \underline{S}imple, \underline{M}ediate, and \underline{H}ard sets if the number of API calls to tackle the instruction is $\le 3$, within $[4, 6]$, and $\ge 6$, respectively. }
\label{tab:model-eval-combined}
\end{table*}



\stitle{Metrics}. 
We evaluate performance using both instruction-level accuracy (\textbf{iACC}) and session-level accuracy (s\textbf{ACC}). In the instruction-level setting, the agent operates on the initial document state associated with each instruction and executes that instruction in isolation. In contrast, session-level evaluation requires the agent to begin from the session’s initial state and sequentially complete all instructions in order. 
In addition, we report the average number of API calls and average token usage required to complete an instruction or an entire session. These metrics serve as indicators of the agent’s automation efficiency.

\stitle{LLMs and generation parameters}. 
We evaluate \modelname using four recent LLM: two open-source models DeepSeek-V3 (used by default) and Qwen-Plus, and two proprietary models GPT-4.1 and Gemini 2.5 Pro. For embedding-based retrieval, we use text-embedding-v4. During all experiments, the temperature is fixed at 0.1, and all other generation parameters are kept at their default values.

We next present our findings.

\eat{ 
We adopt a multi-level evaluation framework based on our RealtimeEvaluator implementation.

Let $\mathcal{T} = \{t_1, \dots, t_n\}$ be tasks ($n = 1708$), $\mathcal{S} = \{s_1, \dots, s_m\}$ be sessions ($m = 250$). For task $t_j$: $r_j \in \{0,1\}$ (success), $a_j \in \mathbb{N}^+$ (API calls), $k_j \in \mathbb{R}^+$ (tokens).

\paragraph{Task-Level Metrics}
\begin{equation}
\label{eq:task-metrics}
\text{TAR} = \frac{1}{|\mathcal{T}|} \sum_{j=1}^{|\mathcal{T}|} r_j, \quad
\text{AAC} = \frac{1}{|\mathcal{T}|} \sum_{j=1}^{|\mathcal{T}|} a_j, \quad
\text{ATC} = \frac{1}{|\mathcal{T}|} \sum_{j=1}^{|\mathcal{T}|} k_j
\end{equation}

\paragraph{Session-Level Metrics}
For session $s_i$ with tasks $\mathcal{T}_i$, session success requires all tasks to succeed: $R_i = \prod_{t_j \in \mathcal{T}_i} r_j$.
\begin{align}
\label{eq:session-metrics}
\text{SCR} &= \frac{1}{|\mathcal{S}|} \sum_{i=1}^{|\mathcal{S}|} R_i, \quad
\text{SAC} = \frac{1}{|\mathcal{S}|} \sum_{i=1}^{|\mathcal{S}|} \sum_{t_j \in \mathcal{T}_i} a_j \\
\text{STC} &= \frac{1}{|\mathcal{S}|} \sum_{i=1}^{|\mathcal{S}|} \sum_{t_j \in \mathcal{T}_i} k_j \nonumber
\end{align}

\paragraph{Global and Difficulty-Stratified Metrics}
\begin{equation}
\label{eq:global-difficulty-metrics}
\text{TAR}_d = \frac{1}{|\mathcal{T}_d|} \sum_{t_j \in \mathcal{T}_d} r_j, \quad
\text{ATC}_d = \frac{1}{|\mathcal{T}_d|} \sum_{t_j \in \mathcal{T}_d} k_j
\end{equation}
where $d \in \{\text{simple}, \text{medium}, \text{difficult}\}$.
}  




\subsection{Overall Performance Comparison}

To demonstrate the effectiveness of \modelname, we compare its performance against three baseline approaches introduced earlier. The results are summarized in Table~\ref{tab:baselines}, and we highlight several key observations:

\begin{itemize}
    \item The retrieval-only method performs poorly, achieving just 14\% iACC and a mere 4.4\% sACC. This confirms that simple semantic retrieval is insufficient for complex, sequential workflows. The method lacks awareness of the document state and cannot dynamically schedule actions, making it incapable of resolving inter-instruction dependencies or disambiguating intent in multi-step tasks.

    \item The reasoning-only method improves over retrieval, reaching approximately 25\% sACC. However, the performance remains limited, highlighting the brittleness of relying solely on in-context LLM reasoning for long-horizon tasks. This baseline struggles with long input contexts (commonly referred to as the ``lost in the middle'' problem) and lacks structured planning or validation, leading to frequent execution failures.

    \item Hybrid is the strongest among the baselines, achieving 64\% iACC but only 35\% sACC. Its primary limitation is that it generates a complete plan upfront and lacks any recovery mechanism. Consequently, a single mid-process error can cascade and derail the entire session, resulting in a significant drop in session-level accuracy.
    
    \item \modelname outperforms all baselines by a large margin on both instruction- and session-level metrics. It achieves over 90\% iACC, demonstrating highly reliable instruction execution. More notably, it reaches a session-level accuracy (sACC) of 62\%, outperforming the next best baseline by a relative gain of 76\%. This improvement is achieved with only 26.5\% more tokens, a modest increase given the substantial accuracy boost. Additionally, AutoDW's stepwise planning does not result in excessive API calls; its API usage remains comparable to that of other reasoning-based methods.
\end{itemize}

These results clearly demonstrate the effectiveness and efficiency of our \modelname framework. The integration of stepwise planning and adaptive rollback successfully maintains long-horizon alignment with user intent and evolving document state, setting a new state-of-the-art for complex document workflow automation.



\eat { 
\begin{table*}[h]
\centering
\setlength{\tabcolsep}{1.2mm}  
\begin{tabular}{l|ccc|ccc|ccc|ccc}
\hline
& \multicolumn{3}{c|}{\textbf{Instruction-level}}  & \multicolumn{3}{c|}{\textbf{Session-level}} 
& \multicolumn{3}{c|}{\textbf{iACC\% by difficulty}} & \multicolumn{3}{c}{\textbf{sACC\% by difficulty}} \\
& \textbf{iACC\%} & \textbf{\#APIs} & \textbf{\#Tokens} & \textbf{sACC\%} & \textbf{\#APIs} & \textbf{\#Tokens} & \textbf{S} & \textbf{M} & \textbf{H} & \textbf{S} & \textbf{M} & \textbf{H} \\
\hline
No-Rollback    & 71.77 & 4.12 & 26.80 & 35.60 & 28.10 & 183.11 & 67.35 & 79.34 & 68.61 & 24.24 & 26.81 & 29.43 \\
Param-Only     & 76.81 & 4.68 & 31.53 & 43.60 & 32.04 & 215.23 & 71.72 & 83.30 & 72.41 & 28.57 & 31.52 & 34.54 \\
\textbf{\modelname} & 90.33 & 5.21 & 42.82 & 62.00 & 34.45 & 284.91 & 94.54 & 90.02 & 86.33 & 39.17 & 43.25 &46.18 \\ 
Double-Round   & 91.42 & 6.95 & 48.60 & 64.80 & 47.52 & 331.71 & 88.25 & 95.87 & 90.13 & 44.21 & 48.63 & 53.00 \\
\hline
\end{tabular}
\caption{Ablation study results across rollback strategies.}
\label{tab:combined-ablation-compact}
\end{table*}
} 

\subsection{Robustness Study}

We next conduct a robustness study to evaluate the generalization ability of \modelname under varying circumstances. Specifically, we examine two scenarios: (1) using different backbone LLMs, and (2) operating on tasks of varying difficulty. To this end, we test \modelname with four LLMs and further stratify the iACC result by difficulty level (\ie easy, medium, and hard) based on the number of APIs per instruction. The results are presented in Table~\ref{tab:model-eval-combined}.

Across all tested LLMs, \modelname demonstrates strong and consistent performance. With Qwen-Plus, it achieves 82.8\% iACC and 53.6\% sACC, lower than other LLMs. This is likely due to its relatively weaker capabilities compared to other tested models.\footnote{\url{https://lmarena.ai/leaderboard}} Nevertheless, it still significantly outperforms the strongest baseline (Hybrid). The other three LLMs yield similar results, achieving over 90\% iACC and 62–67\% sACC, with comparable API calls and token usage.
Interestingly, we observe greater variation across LLMs (excluding Qwen) at the session level (a 5.2\% gap in sACC) than at the instruction level (1.7\% gap in iACC). This suggests that \modelname is capable of leveraging the planning and reasoning strengths of more capable LLMs more effectively in complex, multi-step workflows.

For the fine-grained difficulty-based analysis, we observe the expected trend: accuracy decreases as task difficulty increases. However, \modelname maintains strong performance even on the hard subset. With the three stronger LLMs, it achieves iACC above 86\%, which represents only 4.4\% drops from the overall instruction-level accuracy. This mild degradation demonstrates that \modelname is robust to increasing task complexity, ensuring its practical applicability in real-world usage scenarios.



\begin{figure}[tb]
\centering
\includegraphics[width=\columnwidth]{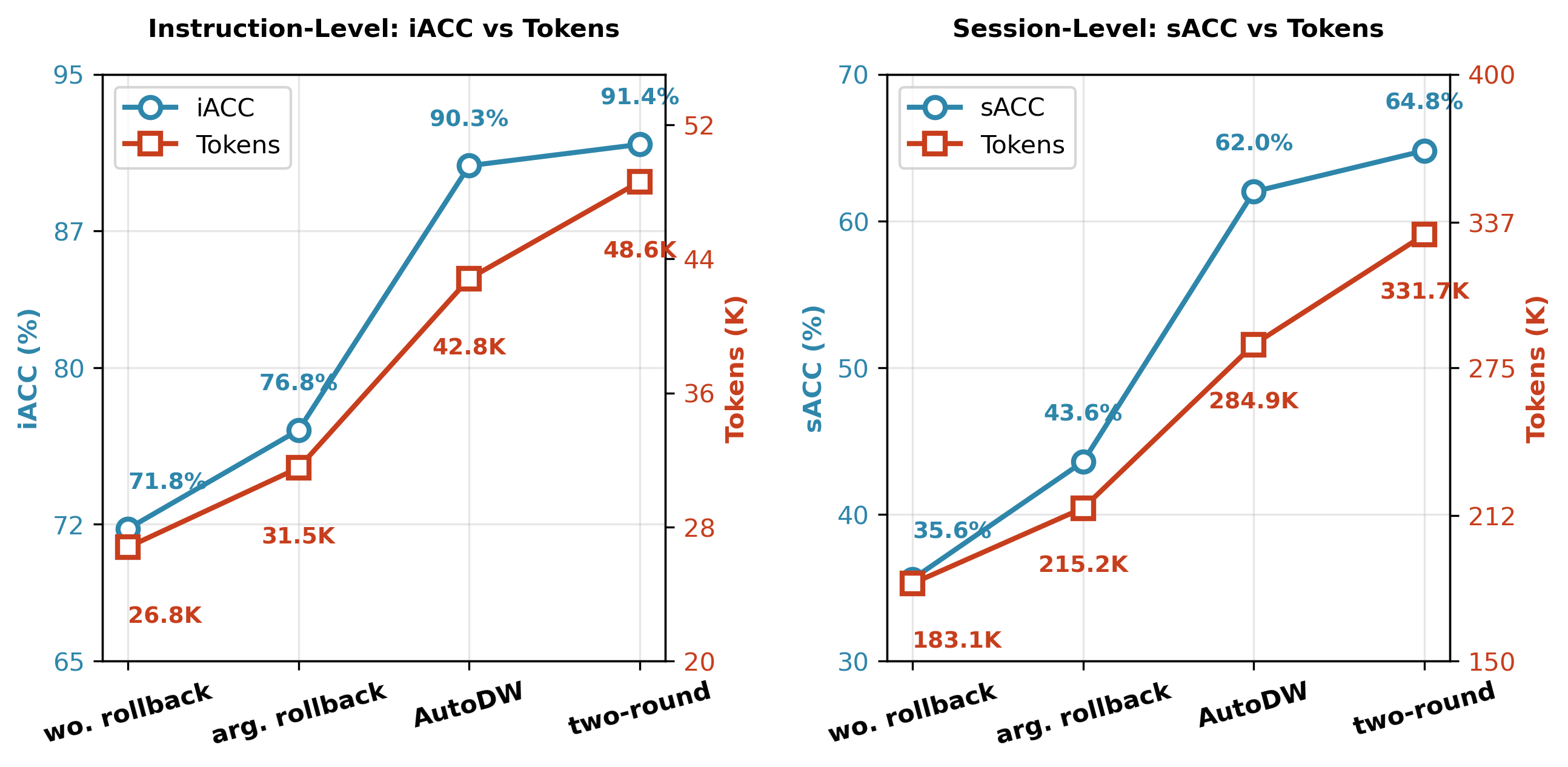}
\caption{Accuracy and token usage of different rollback strategies.}
\label{fig:accuracy_token_analysis}
\end{figure}

\subsection{Ablation Study on Rollback Mechanisms}

We conclude our experimental analysis with an ablation study to validate the design choices underlying \modelname. As discussed earlier, \modelname integrates two key innovations: stepwise planning and adaptive rollback. However, fully ablating stepwise planning is non-trivial, as rollback inherently requires dynamic re-planning. Therefore, in this study, we focus on ablating the rollback mechanism. Specifically, we evaluate three variants of \modelname:
(1) wo. rollback, a version that uses stepwise planning only,
(2) arg. rollback which applies a single round of argument-level rollback per action, and
(3) two-round, which performs two full rounds of rollback if needed.
We compare these variants with the standard \modelname in terms of instruction-level accuracy (iACC), session-level accuracy (sACC), and token usage. The results are summarized in Fig.~\ref{fig:accuracy_token_analysis}.

As shown in the figure, both accuracy and token usage increase as more rollback budget is allocated (from left to right). Specifically, the iACC scores for the four configurations are 71.8\%, 76.8\%, 90.3\%, and 91.4\%, while the corresponding sACC scores are 35.6\%, 43.6\%, 62.0\%, and 64.8\%. Although token usage grows roughly linearly from wo. rollback to two-round, the associated accuracy improvements are non-linear. The most substantial gain occurs between arg. rollback and \modelname, with an improvement of 13.5\% in iACC and 18.4\% in sACC. In contrast, further extending to two-round yields only marginal improvements, \ie an additional 1.1\% iACC and 2.8\% sACC, representing less than 20\% of the previous gain.
These results suggest that AutoDW’s current rollback strategy, \ie a single round of dual-level rollback, achieves an effective balance between performance and token cost. It delivers most of the accuracy benefit (\ie +74\% from 35.6\% to 62\%) while keeping token usage affordable (\ie +56\%), making it a practical design choice for real-world deployment.


\section{Conclusion and Future Work}
\label{sec:conclusion}

In this paper, we presented \modelname, a novel framework that integrates stepwise planning and adaptive dual-level rollback (acting as a dynamic fault-tolerance gate) to automate complex document workflows with robust fault tolerance. $\text{\modelname}$ incrementally plans $\text{API}$ actions conditioned on the evolving document state, enabling the system to detect and promptly correct errors before they cascade. To evaluate this capability, we introduced \modelname, a comprehensive bilingual benchmark of 250 multi-turn sessions with 1,708 human-annotated instructions, designed for rigorous assessment of long-horizon automation where sessions require 15 to 75 API calls. Experimental results demonstrate that $\text{\modelname}$ achieves state-of-the-art performance, outperforming baselines by at least 40\% and 76\% in instruction and session metrics, respectively. Our ablation study critically reveals the adaptive rollback module alone contributes a 74\% relative improvement in session-level success. Looking forward, we envision several promising directions for future research, including exploring more sophisticated planning techniques (e.g., hierarchical or graph-based modeling), extending $\text{\modelname}$ to support a broader range of document types and modalities (e.g., spreadsheets, PDFs), improving rollback efficiency through lightweight validation models, investigating human-in-the-loop strategies to handle instruction ambiguity, and expanding $\text{\benchname}$ to include collaborative and multi-agent scenarios.

\bibliography{aaai2026.bib}


\end{document}